\documentclass[sigconf]{aamas} 

\usepackage{balance} 
\usepackage{subfigure}

\setcopyright{ifaamas}
\acmConference[AAMAS '22]{Proc.\@ of the 21st International Conference
on Autonomous Agents and Multiagent Systems (AAMAS 2022)}{May 9--13, 2022}
{Auckland, New Zealand}{P.~Faliszewski, V.~Mascardi, C.~Pelachaud,
M.E.~Taylor (eds.)}
\copyrightyear{2022}
\acmYear{2022}
\acmDOI{}
\acmPrice{}
\acmISBN{}

\acmSubmissionID{14}

\title[AAMAS-2022 Formatting Instructions]{SIDE: State Inference for Partially Observable Cooperative Multi-Agent Reinforcement Learning}

\author{Zhiwei Xu, Yunpeng Bai, Dapeng Li, Bin Zhang, Guoliang Fan}
\affiliation{
  \institution{Institute of Automation, Chinese Academy of Sciences\\School of Artificial Intelligence, University of Chinese Academy of Sciences}
  \city{Beijing}
  \country{China}}
\email{{xuzhiwei2019, baiyunpeng2020, lidapeng2020, zhangbin2020, guoliang.fan}@ia.ac.cn}

\begin{abstract}
As one of the solutions to the decentralized partially observable Markov decision process (Dec-POMDP) problems, the value decomposition method has achieved significant results recently. However, most value decomposition methods require the fully observable state of the environment during training, but this is not feasible in some scenarios where only incomplete and noisy observations can be obtained. Therefore, we propose a novel value decomposition framework, named State Inference for value DEcomposition (SIDE), which eliminates the need to know the global state by simultaneously seeking solutions to the two problems of optimal control and state inference. SIDE can be extended to any value decomposition method to tackle partially observable problems. By comparing with the performance of different algorithms in StarCraft II micromanagement tasks, we verified that though without accessible states, SIDE can infer the current state that contributes to the reinforcement learning process based on past local observations and even achieve superior results to many baselines in some complex scenarios.
\end{abstract}


\newcommand{\BibTeX}{\rm B\kern-.05em{\sc i\kern-.025em b}\kern-.08em\TeX}

\begin{document}


\pagestyle{fancy}
\fancyhead{}


\maketitle 


\section{Introduction}

Deep reinforcement learning has recently made breakthroughs in complex scenarios such as Atari games~\cite{Mnih2013PlayingAW}, robot control~\cite{OpenAI2019SolvingRC}, and autonomous driving~\cite{Faust2018PRMRLLR}. In the real world, however, due to the complexity of entity attributes and local observability, it is often impossible to obtain an effective representation for the state of the environment, which has a catastrophic impact on reinforcement learning. Different from the single-agent reinforcement learning tasks, there are multiple entities in the multi-agent system (MAS), so whether the state representation is appropriate plays a more important role in multi-agent reinforcement learning problems. One of the traditional methods is to simply stack the local observations of all agents as the current state representation, but the most direct drawback of this is that as the number of agents increases, the dimension of the state representation will also increase dramatically.

On the one hand, to alleviate the partially observable problems, a lot of work has been proposed. Some model-based reinforcement learning algorithms~\cite{Silver2010MonteCarloPI, Somani2013DESPOTOP, Egorov2015DeepRL} use environmental dynamics information to solve the POMDP problems, but transition functions are typically not available in most real tasks. Besides, because the recurrent neural network has the characteristics of integrating historical information, it was introduced into the vanilla reinforcement learning algorithms~\cite{Hausknecht2015DeepRQ, Wang2020RMADDPGFP, Zhu2017OnID}. In this way, the performance has been improved while the algorithm remains model-free, even if the training of the recurrent neural network may require more trajectories of experience. Besides, there are state estimation methods based on variational inference~\cite{Igl2018DeepVR, Huang2020SVQNSV, Han2020VariationalRM} or belief tracking methods based on particle filters~\cite{Karkus2018ParticleFN, jonschkowski18, Ma2020Discriminative}, but these approaches may not be practical in multi-agent cases. On the other hand, some work~\cite{Lample2017PlayingFG, Jaderberg2017ReinforcementLW} promotes the neural network to extract helpful information from the state of the complex environment by adding auxiliary tasks mainly including predicting the state of the next timestep. Intuitively, the key limitation of these studies is that they do not take the cases that the current states are unobserved into account.

As a notorious problem in MAS, Dec-POMDP~\cite{Oliehoek2016ACI} describes some collaboration problems. Since the global reward function is shared, it is necessary to allocate credit to each agent. In recent years, research on value decomposition has become very popular because of its simple implementation and excellent performance. The earliest value decomposition method is Value Decomposition Network (VDN)~\cite{Sunehag2018ValueDecompositionNF} and is followed by QMIX~\cite{Rashid2018QMIXMV}, one of the most popular multi-agent algorithms. Furthermore, many excellent variants of value decomposition methods~\cite{Wang2020QPLEXDD,Mahajan2019MAVENMV,Wang2020ROMAMR,Wang2020RODELR} have been recently proposed. However, most value decomposition methods use the global state by default during centralized training, which is not allowed in some environments where the global state is inaccessible.

In this paper, we propose \textbf{S}tate \textbf{I}nference for value \textbf{DE}composi-tion (SIDE), a state variational estimation framework based on multi-agent value decomposition reinforcement learning methods. SIDE does not require dynamics information of the environment. It uses the variational graph auto-encoder~\cite{Kipf2016VariationalGA} to integrate the local observation of all agents and reconstructs the state while reducing the dimension of the state space. Reinforcement learning and state inference in SIDE are carried out simultaneously, so it can promote the construction of states that is beneficial to maximizing returns. And because SIDE can be seen as a state inference mechanism based on the value decomposition method, it can be applied to any QMIX-style algorithm. We evaluated the performance of SIDE in the StarCraft II Dec-POMDP micromanagement tasks. The experimental results prove that when the true global state is unknown, the performance of SIDE can still be close to or even better than the methods of using states which are set manually.


\section{Related Work}

In this section, we will summarize some recent work on partial observable problems in the case of single-agent and multi-agent respectively.

\noindent
\textbf{Single-Agent}\quad In a POMDP the agent additionally has perceptual uncertainty. Agents usually use past local observations and actions to understand the current state. This understanding is called belief. One of the most popular methods is the belief tracking method~\cite{Karkus2018ParticleFN, jonschkowski18, Ma2020Discriminative} based on particle filters, but this method often requires the preset model and state representation. Some similar work is to integrate information at different timesteps by introducing recurrent neural networks~\cite{Hausknecht2015DeepRQ, Zhu2017OnID}, thereby allowing us to loosen the assumptions of earlier model-based methods and apply a model-free technique instead. The other is the state generation models~\cite{Igl2018DeepVR, Han2020VariationalRM} based on variational inference. The generative model can abstract observations to output an effective state in compact latent spaces by optimizing the evidence lower bound. Sequential variational soft Q-learning network (SVQN)~\cite{Huang2020SVQNSV} represents POMDPs as a probabilistic graphical model (PGM)~\cite{Koller2009ProbabilisticGM} and combines it with variational inference. Unlike the standard variational auto-encoder (VAE)~\cite{Kingma2014AutoEncodingVB}, the distribution of state hidden variables is conditional on previous hidden states, so SVQN is equipped with an additional generative model to learn the conditional prior of hidden states. However, SVQN is only applicable to single-agent environments and there are still some interesting problems to be addressed in multi-agent systems.

\noindent
\textbf{Multi-Agent}\quad In the multi-agent environment, in addition to the partial observability of the environment itself, because each agent sometimes cannot observe other agents, this makes the problem of environmental uncertainty more complicated. Communication~\cite{Sukhbaatar2016LearningMC, Foerster2016LearningTC, Peng2017MultiagentBN, Kim2019LearningTS} can build a channel of information between various agents, thereby alleviating partial observable problems. Moreover, the framework of centralized training with decentralized execution (CTDE) allows the agent to share all the information with other agents during training. Much excellent work~\cite{Lowe2017MultiAgentAF, Foerster2018CounterfactualMP, Iqbal2019ActorAttentionCriticFM} has been proposed based on this paradigm and one of the most representative studies is the value decomposition method. The main idea of QMIX, a popular value decomposition method, is to input the action-values of all agents into the mixing network and then output the global action-value. The parameters of the mixing network are the output of a few hypernetworks~\cite{Ha2017HyperNetworks} whose input is the global state. Although the above methods have effects in the multi-agent partially observable environments, they can only alleviate the problem of incomplete information between agents, rather than the invisible information problem of the environment itself.

Enlightened by SVQN, we have perfected the reconstruction of unobserved information in the multi-agent environment itself, that is, using variational inference with past local observations. Besides, we also confirmed whether QMIX can still outperform VDN under the premise of obtaining the same amount of information as VDN. Finally, it is proved according to the experimental results that in different scenarios, compared with QMIX under different state representation definitions, our proposed SIDE can find the most appropriate state representation. To illustrate it, we carry out the visualization of the embedding of latent states learned with our proposed method. The visualization results intuitively show that the latent state variables learned by SIDE are conducive to reinforcement learning.

\begin{figure*}[tp]
    \subfigure[MDPs]{
        \includegraphics[width=0.27\textwidth]{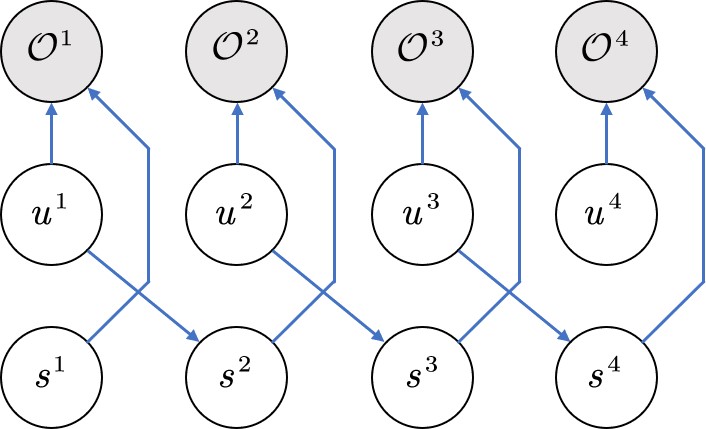}
    }
    \hspace{0.3 in}
    \subfigure[POMDPs]{
        \includegraphics[width=0.27\textwidth]{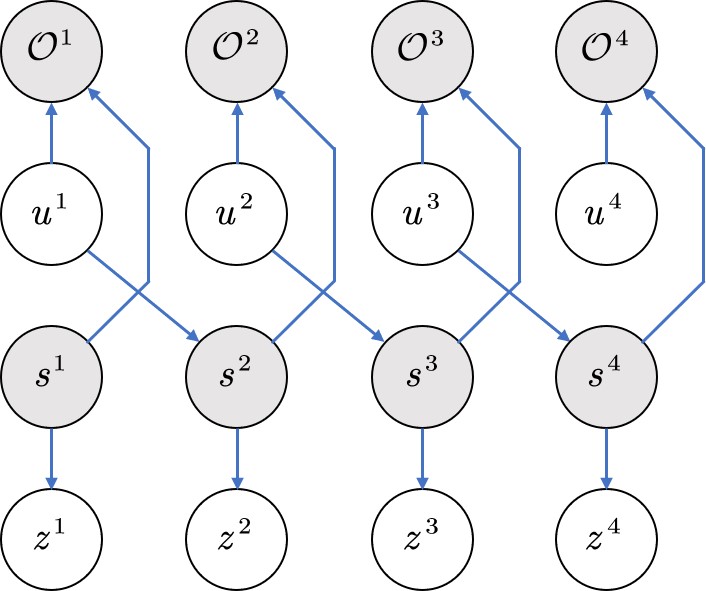}
    }
    \hspace{0.3 in}
    \subfigure[Dec-POMDPs]{
        \includegraphics[width= 0.27\textwidth]{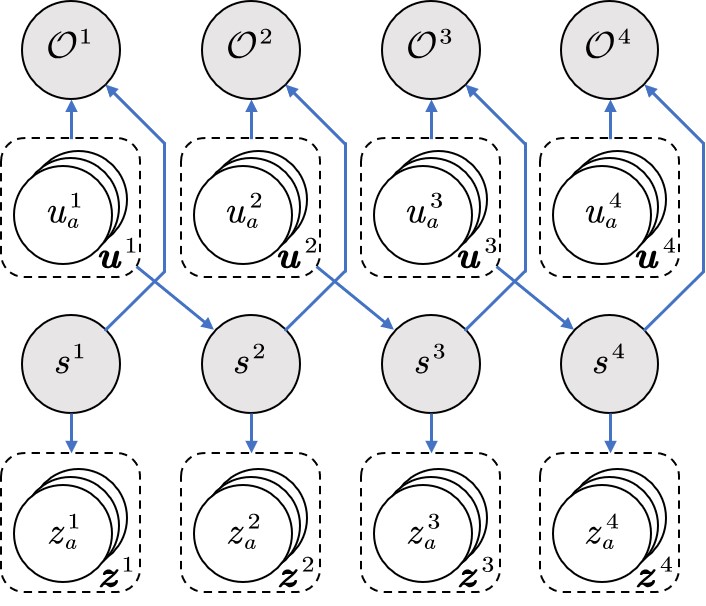}
    }
    \caption{The graphical models for MDPs(a), POMDPs(b) and Dec-POMDPs(c). Grey nodes are hidden and white nodes are observable.}
    \label{fig:pgm}
\end{figure*}

\section{Background}
\subsection{Dec-POMDP}
The Dec-POMDP is defined as a tuple $\langle \mathcal{S},\mathcal{U}, \mathcal{P}, r, \mathcal{Z}, {O}, n, \gamma\rangle$. $s \in \mathcal{S}$ represents the state of the environment. At each timestep, each agent $a \in \mathcal{A} := \left \{1, ..., n\right \}$ will take an action $u_a \in \mathcal{U}$ and then the local observation of each agent $z_a \in \mathcal{Z}$ is obtained by ${O}(s, a): \mathcal{S} \times \mathcal{A} \rightarrow \mathcal{Z}$. The actions of all agents form the joint action $\boldsymbol{u} \in \boldsymbol{\mathcal{U}} \equiv \mathcal{U}^n$. The state transition function, which generates the next state of the environment, is defined as $\mathcal{P}(s'|s, \boldsymbol{u}) : \mathcal{S} \times \boldsymbol{\mathcal{U}} \times \mathcal{S}\rightarrow[0,1]$. In Dec-POMDPs, a common joint reward function $r(s, u): \mathcal{S} \times \boldsymbol{\mathcal{U}} \rightarrow \mathbb{R}$ is provided for all agents. $\gamma \in [0, 1)$ is the discount factor. The goal is to maxmise the discounted return $R^t = \sum_{l=0}^\infty (\gamma)^lr^{t+l}$ in Dec-POMDPs.

In one episode, each agent will get the respective action-observa-tion history $\tau_a \in T \equiv(\mathcal{Z}\times \mathcal{U})$. $\pi_a (u_a | \tau_a) : T \times \mathcal{U} \rightarrow \left[0, 1\right]$ denotes the policy of each agent. Note that for the convenience of presentation in this paper, the superscript represents the time and the subscript represents the identity number of the agent.

\subsection{Value Decomposition}
In the field of cooperative multi-agent reinforcement learning, it is not feasible to train each agent individually or treat all agents as a whole for joint training when the number of agents is large enough. Therefore, some research has proposed various methods between the above two extreme methods such as VDN, QMIX. These value decomposition methods try to achieve automated learning of decomposition of the joint value function based on the Individual-Global-Max (IGM)~\cite{Son2019QTRANLT}, where IGM assumes that the optimality of each agent is consistent with the optimality of all agents. The equation that describes IGM is as follows:
\begin{equation*}
\arg \max _{\boldsymbol{u}} Q_{\mathrm{tot}}(\boldsymbol{\tau}, \boldsymbol{u})=\left(\begin{array}{c}
\arg \max _{u_{1}} Q_{1}\left(\tau_{1}, u_{1}\right) \\
\vdots \\
\arg \max _{u_{n}} Q_{n}\left(\tau_{n}, u_{n}\right)
\end{array}\right),
\end{equation*}
where $\boldsymbol{\tau} \in T^n$ represents the joint action-observation histories of all agents. $Q_{tot}$ is global action-value function and $Q_a$ is the individual ones.

VDN assumes that the joint value function is linearly decomposable. However, the linear assumption is too simple to fit most scenarios. Therefore, QMIX introduces a nonlinear global value function and assumes that the joint action-value function $Q_{tot}$ is monotonic to the individual action-value function $Q_a$ in order to satisfy the IGM assumption. In addition, many studies have presented interesting varieties of value decomposition and focused on different problems.

\subsection{Probabilistic Graphical Models}
The probabilistic graphical model (PGM)~\cite{Larraaga2002AnIT} is studied on the basis of probability theory and graph theory. It visualizes the probability model through the structure of the graph, allowing us to understand the relationship between variables in a complex distribution. In ~\cite{Levine2018ReinforcementLA}, the MDP problem is embedded in the PGM framework, so that the reinforcement learning problems can be viewed from another perspective. According to the paradigm of PGM, given the state $s\in \mathcal{S}$ and the action $u \in \mathcal{U}$, ~\cite{Levine2018ReinforcementLA} introduced the optimality variable $\mathcal{O}$ to solve the optimality of MDP control problems. This variable $\mathcal{O}$ is a binary random variable. When $\mathcal{O}^t=1$, it means that the system at timestep $t$ is optimal. On the contrary, when $\mathcal{O}^t=0$ means it is not optimal. Define the distribution of the variable $\mathcal{O}$ as:
\begin{equation*}
    p\left(\mathcal{O}^{t}=1 \mid s^{t}, u^{t}\right) \propto \exp \left(r\left(s^{t}, {u}^{t}\right)\right).
\end{equation*}
Then its variational evidence lower bound can be obtained as:
\begin{equation}
\begin{aligned}
    \log p&\left(\mathcal{O}^{1: T}\right) \geq \\
    &\mathbb{E}_{\left(s^{1: T}, u^{1: T}\right) \sim \pi\left(s^{1: T}, u^{1: T}\right)}\left[\sum_{t=1}^{T} r\left(s^{t}, u^{t}\right)-\log \pi\left(u^{t} \mid s^{t}\right)\right],
\end{aligned}
\label{eq:mdp_elbo}
\end{equation}
where $\pi(u^t\mid s^t)$ is the policy function. Therefore, increasing the optimal probability of the entire MDP $p(\mathcal{O}^{1:T})$ is to optimize its optimal lower bound, which leads to the maximum entropy reinforcement learning~\cite{Haarnoja2017ReinforcementLW}. Figure~\ref{fig:pgm}(a) shows the PGM of the ordinary MDP.

SVQN combines latent state inference with maximum entropy reinforcement learning under a probabilistic graphical model, and optimizes these two modules simultaneously to address partial observation problems. Since the state $s$ cannot be observed in the POMDP, it needs to be inferred from the local observation $z$ and the action $u$. Such cases are depicted in Figure~\ref{fig:pgm}(b). SVQN proposes a variational evidence lower bound that is different from the Equation~(\ref{eq:mdp_elbo}):

\begin{align}
&\log p\left(\mathcal{O}^{0: T}, u^{0: T}, z^{1: T}\right) \nonumber\\
=&\log \mathbb{E}_{q_{\theta}\left(s^{1: T} \mid \mathcal{O}^{1: T}, u^{0: T}, z^{1: T}\right)}\left[\frac{p\left(s^{1: T}, \mathcal{O}^{0: T}, u^{0: T}, z^{1: T}\right)}{q_{\theta}\left(s^{1: T} \mid \mathcal{O}^{0: T}, u^{0: T}, z^{1: T}\right)}\right] \nonumber\\
\geq& \mathbb{E}_{q_{\theta}\left(s^{1: T} \mid \mathcal{O}^{1: T}, u^{0: T}, z^{1: T}\right)} \log \left[\frac{p\left(s^{1: T}, \mathcal{O}^{0: T}, u^{0: T}, z^{1: T}\right)}{q_{\theta}\left(s^{1: T} \mid \mathcal{O}^{0: T}, u^{0: T}, z^{1: T}\right)}\right]\nonumber\\
=& \mathbb{E}_{q_{\theta}\left(s^{1: T} \mid \mathcal{O}^{1: T}, u^{0: T}, z^{1: T}\right)} \sum_{t=1}^{T}\left\{r\left(s^{t}, u^{t}\right)+\log \left[p\left(u^{t}\right) p\left(z^{t} \mid s^{t}\right)\right]\right. \nonumber\\
&  \left.-\mathcal{D}_{K L}\left[q_{\theta}\left(s^{t} \mid s^{t-1}, u^{t-1}, z^{t}\right) \| p\left(s^{t} \mid s^{t-1}, u^{t-1}\right)\right]\right\},
\label{eq:sqvn_elbo}
\end{align}
where $p(u^t)$ is the action prior, for simplicity it is often regarded as a uniform distribution and it has nothing to do with the optimization of ELBO. And $\theta$ is the parameter of the approximate function $q(\cdot)$. $p(z^t \mid s^t)$ implies that the state $s^t$ can generate the current observation $z^t$. For the last term KL divergence, since $p(s^t \mid s^{t-1},u^{t-1})$ is unknown, SVQN presents an additional generative model $p_{\theta}^{ \text {prior }}\left(s^{t} \mid s^{t-1}, u^{t-1}\right)$ to approach the real distribution $p(s^t \mid s^{t-1}, u^{t-1})$. The structured variational inference is used to optimize the ELBO, that is, SVQN solves the two problems of optimal control and state inference by jointly optimizing the loss functions of two variational auto-encoders and the soft Q-learning.

\begin{figure*}[t]
    \centering
    \includegraphics[width=0.8\textwidth]{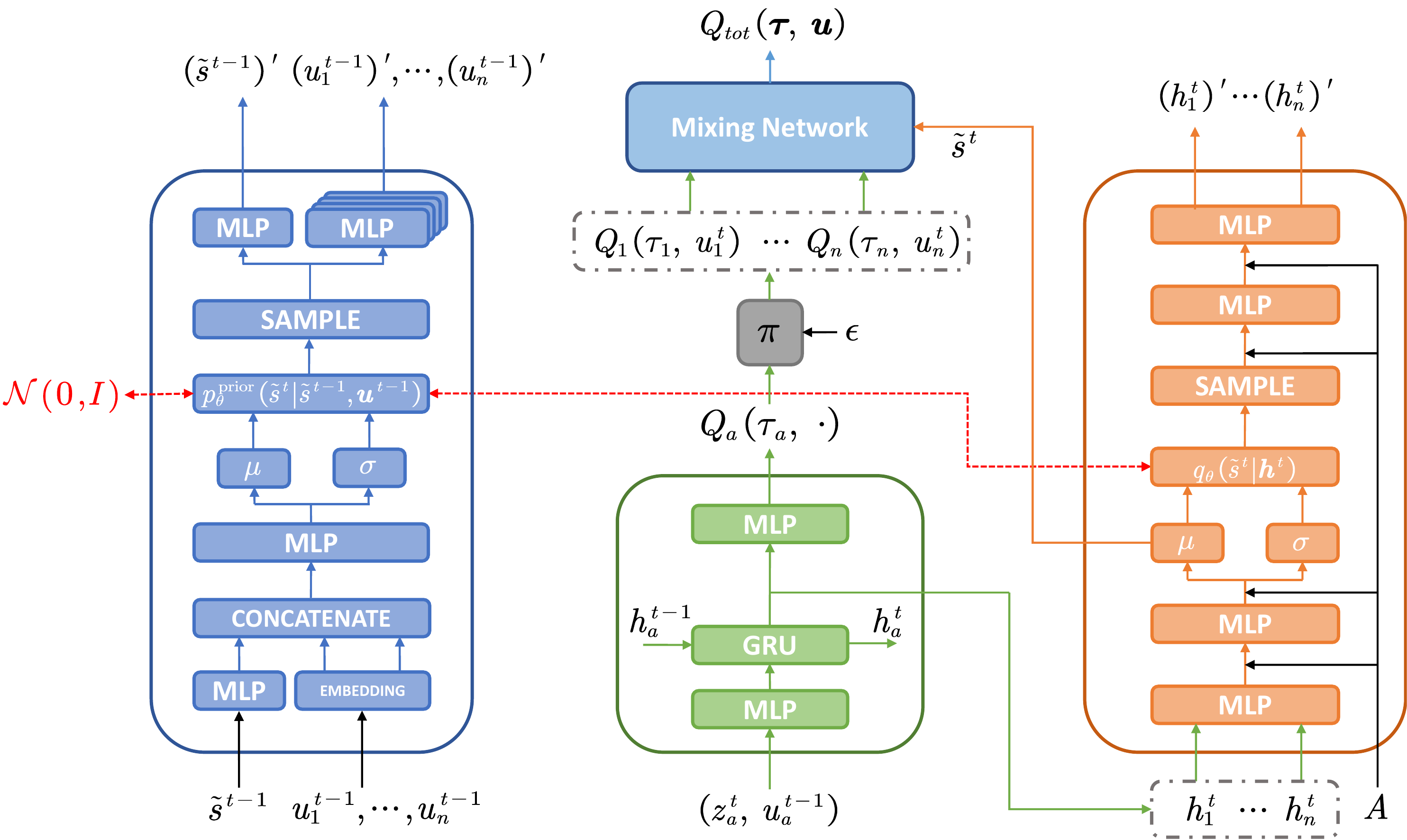}
    \caption{The overall architecture of SIDE. The left part is the prior function $p_\theta^\text{prior}(\cdot)$ and the state inference function $q_\theta(\cdot)$ based on the VGAE is on the right. The red dotted line connects the two distributions that need to calculate the KL divergence.}
    \label{fig:Framework}
\end{figure*}

\subsection{Variational Graph Auto-Encoders}
The variational graph auto-encoder (VGAE), a framework that combines both variational auto-encoders and graph networks, is increasingly being used in graph structure data. Given a graph $\mathcal{G}$, its adjacency matrix is $\mathbf{A}$, and the node feature matrix is $\mathbf{X}$. The simplest VGAE is composed of a graph convolutional network (GCN)~\cite{Defferrard2016ConvolutionalNN} encoder and an inner product decoder. First, each node $i$ is mapped to a random variable distribution $q(\mathbf{z_i} \mid \mathbf{X}, \mathbf{A})$ through GCN. After that, the latent variable representation $\mathbf{z}_i$ of each node is obtained by the reparameterization trick~\cite{Kingma2014AutoEncodingVB}. The decoder treats the inner product $\mathbf{A}^\prime=\sigma(\mathbf{Z}\mathbf{Z}^T)$ with respect to the hidden variables of all nodes as the reconstructed adjacency matrix, where $\sigma(\cdot)$ generally indicates the logistic sigmoid function. The optimization goal of VGAE is to minimize the difference between the real adjacency matrix and the reconstructed one, as well as the KL divergence between the random latent variable distribution and the prior distribution $p(\mathbf{Z})$ (usually set to a standard Gaussian distribution). The loss function can be written as:
\begin{equation*}
    \mathcal{L}=\mathbb{E}_{q(\mathbf{Z} \mid \mathbf{X}, \mathbf{A})}[\log p(\mathbf{A}^\prime \mid \mathbf{Z})]-\operatorname{KL}[q(\mathbf{Z} \mid \mathbf{X}, \mathbf{A}) \| p(\mathbf{Z})].
\end{equation*}
Besides, many variations~\cite{Pan2018AdversariallyRG, Salha2019KeepIS} of the variational graph auto-encoder have been proposed recently. In order to reconstruct the node features $\mathbf{X}$ instead of the adjacency matrix $\mathbf{A}$, the decoder in Graph convolutional Autoencoder using LAplacian smoothing and sharpening (GALA)~\cite{Park2019SymmetricGC} also uses a GCN structure that is symmetrical to the encoder. Contrary to the encoder based on Laplacian smoothing, the decoder is built on the basis of Laplacian sharpening. Compared with other variational graph auto-encoder variants, GALA shows superb results. In this paper, we will use GALA to integrate the local observations of all agents to reconstruct the state.


\section{State Inference for Value Decomposition}
In this section, we will elaborate on the proposed algorithm State Inference for value DEcomposition (SIDE). At first we embed the Dec-POMDP into the PGMs and obtain the corresponding evidence lower bound, and then build a specific neural network framework.

\subsection{Variational Lower Bound For Dec-POMDPS}
Similar to the single-agent POMDP problem, in Dec-POMDP, the optimal variable $\mathcal{O}$ and the global state $s$ are both hidden variables that cannot be observed. However, the actions and observation variables of each agent in Dec-POMDP are different. Since the reward function is shared by all agents and $\mathcal{O}$ is related to the reward $r$, all agents also share an optimal variable $\mathcal{O}$ in PGMs . The PGMs framework of Dec-POMDP is shown in Figure~\ref{fig:pgm}(c).

First, we construct an inference function $q_\theta(\cdot)$ for the latent state, where $\theta$ is the learnable parameter. Since we need to solve the two problems of optimal control and state inference at the same time, here we use structured variational inference to optimize the variational evidence lower bound of Dec-POMDP, where the optimal policy we use the function $q_\pi(\cdot)$ to approach. It should be noted that, in order to facilitate comparison, we use the vanilla q-learning method instead of the soft q-learning method to train $q_\pi(\cdot)$. From above we can deduce the evidence lower bound of Dec-POMDP:
\begin{align}
&\log p\left(\mathcal{O}^{0: T}, \boldsymbol{u}^{0: T}, \boldsymbol{z}^{1: T}\right) \nonumber\\
=&\log \mathbb{E}_{q_{\theta}\left(s^{1: T} \mid \mathcal{O}^{1: T}, \boldsymbol{u}^{0: T}, \boldsymbol{z}^{1: T}\right)}\left[\frac{p\left(s^{1: T}, \mathcal{O}^{0: T}, \boldsymbol{u}^{0: T}, \boldsymbol{z}^{1: T}\right)}{q_{\theta}\left(s^{1: T} \mid \mathcal{O}^{0: T}, \boldsymbol{u}^{0: T}, \boldsymbol{z}^{1: T}\right)}\right] \nonumber\\
\geq& \mathbb{E}_{q_{\theta}\left(s^{1: T} \mid \mathcal{O}^{1: T}, \boldsymbol{u}^{0: T}, \boldsymbol{z}^{1: T}\right)} \log \left[\frac{p\left(s^{1: T}, \mathcal{O}^{0: T}, \boldsymbol{u}^{0: T}, \boldsymbol{z}^{1: T}\right)}{q_{\theta}\left(s^{1: T} \mid \mathcal{O}^{0: T}, \boldsymbol{u}^{0: T}, \boldsymbol{z}^{1: T}\right)}\right]\nonumber\\
=&\sum_{t=1}^{T}\left\{r\left(s^{t}, \boldsymbol{u}^{t}\right)+\sum_{a=1}^n\log \left[p\left(u_a^{t}\right)\right] + \sum_{a=1}^n\log \left[p\left(z_a^{t} \mid s^{t}\right)\right]\right.\displaybreak\nonumber\\
&-\left.\mathcal{D}_{K L}\left[q_{\theta}\left(s^{t} \mid s^{t-1}, \boldsymbol{u}^{t-1}, \boldsymbol{z}^{t}\right) \| p\left(s^{t} \mid s^{t-1}, \boldsymbol{u}^{t-1}\right)\right]\right\},
\label{eq:SIDE_elbo}
\end{align}
where $ s^{1: T} \sim q_{\theta}\left(s^{1: T} \mid \mathcal{O}^{1: T}, \boldsymbol{u}^{0: T}, \boldsymbol{z}^{1: T}\right)$. Equation~(\ref{eq:SIDE_elbo}) holds under the condition that the action priors and observations of all agents are conditionally independent of each other given the state:
\begin{equation*}
\begin{aligned}
    p(\boldsymbol{u^t}) &= \prod_{a = 1}^n p(u_a^t), \\
    p(\boldsymbol{z}^t\mid s^t) &= \prod_{a=1}^n p(z_a^t\mid s^t).
\end{aligned}
\end{equation*}

In order to maximize the evidence lower bound obtained, it is necessary to analyze each item in Equation~(\ref{eq:SIDE_elbo}). The first term $r(s^t,\boldsymbol{u}^t)$ can be maximised by reinforcement learning. $p(u_a^t)$ is an action prior, which is a constant. $p(z_a^t\mid s^t)$ represents that the latent state needs to be able to generate local observations of each agent. The negative KL divergence means we need to reduce the KL distance between the inference function $q_{\theta} \left (s ^ {t} \mid s ^ {t-1}, \boldsymbol {u} ^ {t-1}, \boldsymbol {z} ^ {t} \right) $ and the prior $p  \left (s ^ {t} \mid s ^ {t-1}, \boldsymbol {u} ^ {t-1} \right) $. Since $p\left(s^{t} \mid s^{t-1}, \boldsymbol{u}^{t-1}\right)$ is unavailable, we adopt the method similar to SVQN by introducing an additional generative function $p_\theta ^  \text {prior} (s ^ t \mid s ^ {t-1}, \boldsymbol {u} ^ {t-1}) $ to approximate the prior $p (s ^ t \mid s ^ {t-1}, \boldsymbol {u} ^ {t-1}) $. In the next section, we will construct the specific neural network framework of SIDE to optimize the ELBO.

\subsection{Framework}
SIDE is based on the value decomposition methods, so the neural network architecture that implements the part of $q_\pi(\cdot)$ is consistent with the popular value decomposition methods. Each agent corresponds to a DRQN~\cite{Hausknecht2015DeepRQ} as the agent network, whose input is the local observation of each agent and outputs the individual action-value $Q_a$. Several hypernetworks generate the parameters of the mixing network according to the state $s$. Finally, the mixing network merges all individual action-values to obtain a global action-value $Q_{tot}$. Through the above architecture diagram, $q_\pi(\cdot)$ is optimized by the value-based reinforcement learning methods. The only difference is that when the global state is not available, the parameters of the mixing network are determined by the latent state $\tilde{s}$ derived from past information.

Next, we will introduce the two functions $p_\theta^\text{prior}(s^t\mid s^{t-1},\boldsymbol{u}^{t-1})$ and $q_\theta(s^t\mid s^{t-1},\boldsymbol{u}^{t-1},\boldsymbol{z}^t)$. First of all, to approximate the unknown prior distribution $p(s^t\mid s^{t-1},\boldsymbol{u}^{t-1})$, we need to construct a fitting function $p_\theta^\text{prior}(s^t\mid s^{t-1},\boldsymbol{u}^{t-1})$, here we use a standard VAE to implement. The VAE deduces the prior distribution of the current state $s ^ t $ according to the past state $s^{t-1}$ and the past actions $\boldsymbol{u}^{t-1}$ of all agents. However, since the real state $s ^ {t-1} $ is unavailable, we use the inferred latent state $\tilde{s}^{t-1}$. Besides, if the agent's action takes the form of one-hot in the discrete case, then $\boldsymbol{u}^{t-1}$ will be very sparse, especially in the case of a large number of agents. Therefore, we design an embedding layer in the encoder and build a separate Multi-Layer Perceptron (MLP) for each agent in the decoder to output the reconstructed action to alleviate this problem. We think it can accelerate the training. Through the reparameterization trick, the prior distribution of the latent state $p_\theta^\text{prior}(\tilde{s}^t\mid \tilde{s}^{t-1},\boldsymbol{u}^{t-1})$ is easy to obtain.

Similarly, the state inference function $q_\theta(\cdot)$ is also constructed through generative models. In order to reduce the computational complexity, we have done two aspects of work. On the one hand, we regard the hidden output $h_a^t$ of the recurrent neural network in the agent network corresponding to each agent as the integration of all past information of the corresponding individual agent, and assume that the hidden outputs of all agents $\boldsymbol {h}^{t}$ contains all past information of the environment. Then the state inference function $q_\theta(s^t\mid s^{t-1},\boldsymbol{u}^{t-1},\boldsymbol{z}^{t-1})$ can be rewritten as $q_\theta(s^t\mid \boldsymbol{h}^{t})$. It should be noted that in order to stabilize the training process, we use the hidden outputs of the recurrent neural network generated by the target agent networks whose updates are relatively slow. On the other hand, instead of using a fully connected VAE, we use GALA, a variational graph auto-encoder with a symmetrical pair of the encoder and the decoder. One advantage of the graph network is that no matter how the number of agents changes, as long as the feature dimension of the node does not change, then the number of parameters that need to be trained will not change. Another advantage of introducing graph networks is that the relationship between nodes is taken into consideration. The relationship between nodes is represented by the adjacency matrix $A$. Last but not least, the inferred hidden state $\tilde{s}^{t}$ is derived directly from the mean $\mu_\theta$ of the distribution $q_\theta(s^t\mid \boldsymbol{h}^{t})$, instead of sampling from $q_\theta(s^t\mid \boldsymbol{h}^{t})$. That is, the hidden state $\tilde{s}^t$ is obtained by concatenating the latent mean values of all nodes. 

The framework of SIDE is shown in Figure~\ref{fig:Framework}. It is worth mentioning that the mixing network in SIDE can use different forms, so SIDE can be extended to various value decomposition methods.

\begin{figure*}[ht]
    \centering
    \includegraphics[width= 0.95 \textwidth]{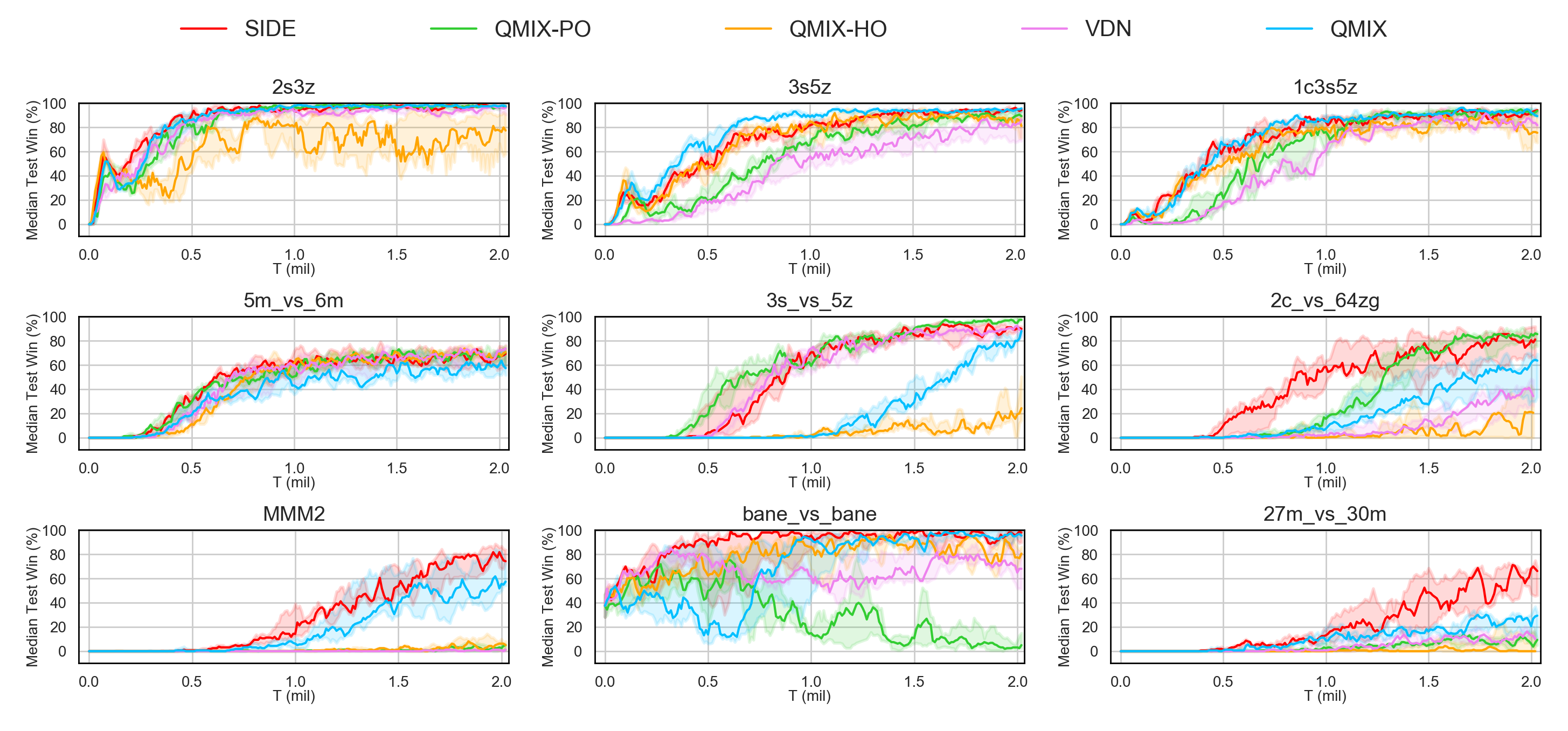}
    \caption{Overall results in different scenarios. Note that none of the algorithms except QMIX can get the true global state.}
    \label{fig:result}
\end{figure*}

\subsection{Loss Function}
In the above framework, a reinforcement learning process and two generative model optimization processes are included. To achieve the two purposes of optimal control planning and global state inference at the same time, the loss functions of the above two processes can be jointly optimized.

For the reinforcement learning process, we use the loss function of QMIX in this paper for simplicity, even though SIDE can be applied to various value decomposition multi-agent reinforcement learning algorithms. We express the parameters of the policy function $p_\pi(\cdot)$ as $\theta_R$. The loss function of the reinforcement learning process can be computed by the following equation:
\begin{equation*}
    \mathcal{L}_{TD}(\theta_R)=\left(y_{tot} - Q_{tot}(\boldsymbol{\tau},\boldsymbol{u}\mid \theta_R)\right)^2,
\end{equation*}
where $y_{tot}$ is the target joint value function and $y_{tot}=r+\gamma \max_{\boldsymbol{u}^\prime}$\\$Q_{tot}(\boldsymbol{\tau}^\prime,\boldsymbol{u}^\prime\mid \theta_R^-)$. $\theta_R^-$ is the parameters of the target network.

For prior model $p_{\theta_I}^\text{prior}(\tilde{s}^t\mid \tilde{s}^{t-1},\boldsymbol{u}^{t-1})$), we utilize the standard VAE training process, that is, to minimize the KL divergence between distribution $p_{\theta_I}^\text{prior}(\tilde{s}^t\mid \tilde{s}^{t-1},\boldsymbol{u}^{t-1})$ and $p(s^t)$, and the difference between the past information $\left((\tilde{s}^{t-1})^\prime, (\boldsymbol{u}^{t-1})^\prime \right)$ obtained by reconstruction and the actual past information $\left(\tilde{s}^{t-1},\boldsymbol{u}^{t-1} \right)$. The two loss functions can be expressed as follows:
\begin{equation*}
    \begin{split}
        &\mathcal{L}^\text{prior}_{KL}(\theta_I) = -\mathcal{D}_{KL}\left[p_{\theta_I}^\text{prior}\left(\tilde{s}^t\mid \tilde{s}^{t-1}, \boldsymbol{u}^{t-1}\right)\| p(s^t)\right],\\
&\mathcal{L}_{RECON}^\text{prior}(\theta_I)=\\
&MSE\left(\tilde{s}^{t-1},(\tilde{s}^{t-1})^\prime\right)+CROSS\_ENTROPY\left(\boldsymbol{u}^{t-1},(\boldsymbol{u}^{t-1})^\prime\right),
    \end{split}
\end{equation*}
where $p(s^t)$ is the standard Gaussian distribution $\mathcal{N}(0, I)$. $MSE(\cdot)$ and $CROSS\_ENTROPY(\cdot)$ represent the mean square error function and the cross entropy loss function respectively.

Finally, on the basis of the loss functions of GALA, The two equations that describes the training loss function of the state inference model $q_{\theta_I}(\tilde{s}^t\mid \boldsymbol{h}^{t})$ are as follows:
\begin{equation*}
\begin{split}
&\mathcal{L}_{KL}(\theta_I) = -\mathcal{D}_{KL}\left[q_{\theta_I}\left(\tilde{s}^{t}\mid \boldsymbol{h}^{t}\right)\| p_{\theta_I}^\text{prior}(\tilde{s}^{t}\mid \tilde{s}^{t-1},\boldsymbol{u}^{t-1})\right],\\
&\mathcal{L}_{RECON}(\theta_I)=MSE\left(\boldsymbol{h}^{t},(\boldsymbol{h}^t)^\prime\right),
\end{split}
\end{equation*}
where $(\boldsymbol{h}^{t})^\prime$ is the reconstructed hidden intermediate variables of agent networks and $\theta_I$ is the parameters of the inferred neural networks.

To sum up, we can get the loss function of the whole framework:
\begin{equation*}
    \mathcal{L}(\theta) = \mathcal{L}_{TD}+\mathcal{L}_{KL}^\text{prior}+\mathcal{L}_{RECON}^\text{prior}+\mathcal{L}_{KL}+\mathcal{L}_{RECON}.
\end{equation*}
By jointly optimizing the loss function, SIDE can infer the latent state that is most conducive to maximizing returns when the global state is unavailable, and gradually achieve optimal control.


\section{Experiments}
We evaluate our proposed algorithm SIDE on the SMAC~\cite{Samvelyan2019TheSM} platform based on StarCraft II. SMAC is a multi-agent testbed dedicated to solving Dec-POMDP problems. Different scenarios correspond to different problems, including heterogeneity, large action spaces, or a large number of agents. Although each agent can only obtain corresponding local observations, the global state can still be obtained by calling the interface of the SMAC. Many value decomposition methods use the global state for training. Even though this does not violate the paradigm of centralized training with decentralized execution, the global state is not available in some environments. 

In this section, we first compare the performance of the three algorithms VDN, QMIX, and SIDE when the state information is not given, and then compare the results with the performance of QMIX in the environment where global state information is available. In this way, it is judged whether SIDE can reconstruct valid state information based on past information.

Furthermore, we carry out the ablation studies to understand the contribution of all components of SIDE. In this paper, we focus on the contribution of the variational graph auto-encoders and the additional prior function. We define the latent state distribution dimension in the variational graph auto-encoder, also called the latent dimension, as $N$. To test whether the latent dimension has a significant influence on the performance of SIDE, we change the value of $N$ in some scenarios and compare the experimental results.

\begin{figure*}[ht]
    \centering
    \includegraphics[width= 0.95 \textwidth]{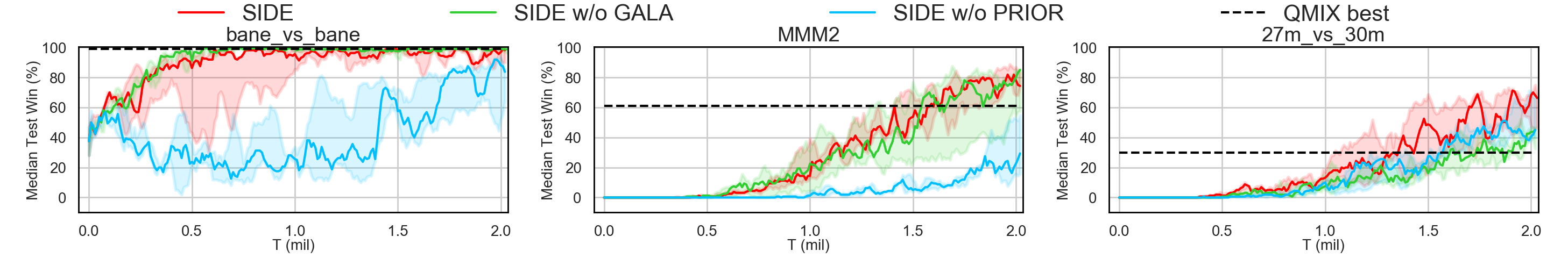}
    \caption{Win rates for SIDE and ablations in three representative scenarios.}
    \label{fig:ablation}
\end{figure*}

\begin{figure*}[ht]
    \centering
    \includegraphics[width= 0.95 \textwidth]{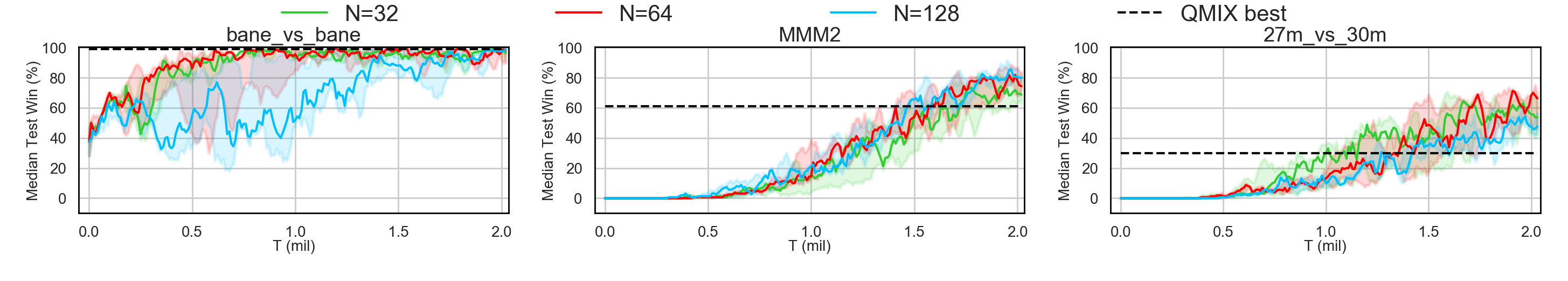}
    \caption{Influence of the different latent dimensions $N$ for SIDE. }
    \label{fig:latent}
\end{figure*}

\subsection{Settings}
The implementations of all algorithms are based on Pymarl, a multi-agent value decomposition algorithm integration platform. Both VDN and QMIX use the default hyperparameter settings. In addition, under the assumption that the perfect state information cannot be obtained, we concatenate the local observations or the hidden intermediate variables of all our own agents and replace the real state with this. We call these two variants QMIX-PO and QMIX-HO, respectively. For ease of comparison between the other baselines, SIDE is implemented on the basis of QMIX. The input adjacency matrix $A$ of the variational graph auto-encoder is set to:
\begin{equation*}
    A_{ij}=\begin{cases}1, &\text{if } i\text{ is alive and } j \text{ is alive}\\0, &\text{otherwise}\end{cases}, \quad \forall A_{ij} \in A.
\end{equation*}
VDN, QMIX-PO, QMIX-HO and SIDE have the same amount of information. By comparing their performance, we can prove that SIDE has an efficient solution to state inference. The above algorithms are also compared with vanilla QMIX to analyze the impact of the lack of global state information.

For ablation studies, we propose two alterations of SIDE. The first ablation replaces the graph auto-encoder in the original SIDE with the fully connected auto-encoder to test the importance of considering the relationship information between agents. And the second one is to remove the additional prior function $p_\theta^\text{prior}(\cdot)$ in the SIDE and directly use only the graph auto-encoder $q_\theta(\cdot)$ to infer the state. To sum up, we change the above two important components of SIDE while keeping other parts unchanged, and test the obtained two ablations in three representative scenarios which are hard or super hard. What's more, we also carry out experiments to test the influence of different latent dimensions $N$.

In this paper, each algorithm runs five experiments independently with different random seeds to avoid the influence of outliers and we evaluate the algorithms every 10,000 time steps. The version of StarCraft II is 4.6.2 (B69232). In our experiments, the versions of GPU and CPU are Nvidia GeForce RTX 3090 and Intel(R) Xeon(R) Platinum 8280 respectively.

\subsection{Validation}

We get the results of the above five algorithms through experiments. In Figure~\ref{fig:result}, the performances of SIDE and the other four algorithms in different scenes are presented. The solid line represents the median win rate, and the 25-75\% percentiles are shown. First, we observe the impact of the lack of information, that is, compare the results of the algorithms QMIX, QMIX-PO and QMIX-HO. In some scenarios, such as \emph{3s5z}, \emph{1c3s5z}, \emph{MMM2}, \emph{bane\_vs\_bane}, and \emph{27m\_vs\_30m}, QMIX-PO performs significantly worse than QMIX. In addition, in the two scenes \emph{5m\_vs\_6m} and \emph{3s\_vs\_5z}, the performance of QMIX-PO and VDN are similar, and both are better than QMIX. This is in line with our intuition. For QMIX-HO, it performs well only in a few scenarios, including \emph{3s5z}, \emph{1c3s5z} and \emph{bane\_vs\_bane}. In the \emph{2c\_vs\_64zg} scenario, the result is more unexpected, because QMIX-PO is far better than QMIX. We think this is because there are too many enemy units, and only focusing on the information of your units significantly improves the learning speed. The above results show that the definitions of state representations that are beneficial to learning in different scenarios are different: in some scenarios agents need global information, while in some scenarios agents tend to focus only on their own information.

From the results, we can easily see that the performance of SIDE in most scenarios close to the best one of the above four algorithms. It is also worth mentioning that SIDE performs far better than other algorithms in some super hard scenarios. This indicates that in different scenarios, SIDE can infer and reconstruct the latent state that is most helpful to the reinforcement learning process, regardless of whether the best state representation contains only its own information, or contains unobserved information, or even other unknown forms. Therefore SIDE does not need the true global state at all. Table~\ref{tab:result} illustrates the median test win rate of different algorithms. The best performances of the above algorithms are bold and the second-best ones are underlined. Besides, we also compare SIDE with current state-of-the-art value decomposition methods and the results can be found in the appendix. In some complex scenarios, SIDE even outperforms all other popular baselines.

As can be seen from Figure~\ref{fig:ablation}, the information of the relationship between agents plays an important role in some scenarios such like \emph{MMM2} and \emph{27m\_vs\_30m}, but is redundant in few maps like \emph{bane\_vs\_bane} where group relations are not valued. On the whole, we can infer that the graph auto-encoder has contributed to the outstanding performance of SIDE. As the another important component, the prior function significantly improve the learning speed of SIDE, which can be known from the poor performance of SIDE without the prior function. Last, we discuss the influence of different values of $N$. As depicted in Figure~\ref{fig:latent}, all choices of $N$ in the figure can typically generate satisfactory results, which means they all significantly outperform QMIX-PO and QMIX-HO. In practice, we set the latent dimension $N$ to 64.

\begin{figure*}[h]
    \centering
    \includegraphics[width= 0.98 \textwidth]{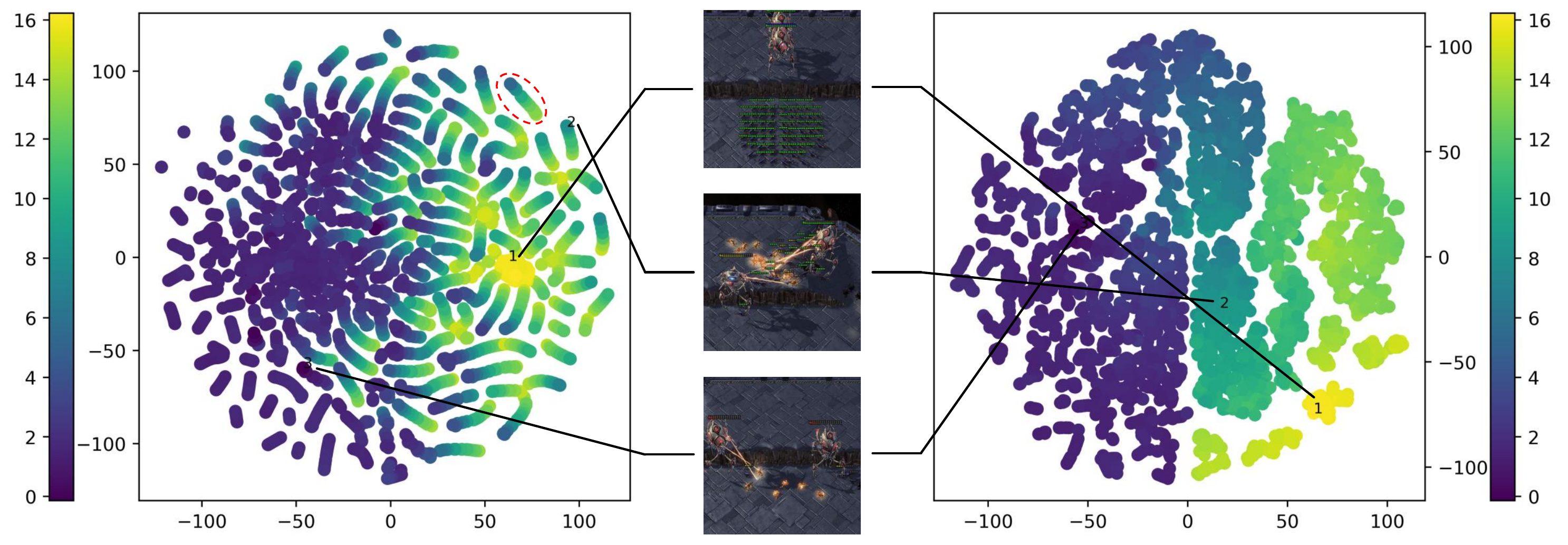}
    \caption{The 2D t-SNE embedding of true states on left, and latent states learned with SIDE on right. Each point is colored according to the discounted return. Three points corresponding to three states in different episodes are selected as examples. We find that the points in the red dashed circle almost belong to the state set of the same episode.}
    \label{fig:vis}
\end{figure*}

\begin{table}[htbp]
\caption{Median performance of the test win ratio (\%) in different scenarios.}
\begin{tabular}{l|cccc|c}
\hline
scenario       & SIDE                & QMIX-PO             & QMIX-HO            & VDN                 & QMIX                \\ \hline
2s3z           & \textbf{99}         & {\underline{98}}    & 81                 & 97                  & {\underline{98}}    \\
3s5z           & \textbf{96}         & 90                  & 93                 & 86                  & {\underline{95}}    \\
1c3s5z         & {\underline{95}}    & {\underline{95}}    & 90                 & 91                  & \textbf{96}         \\
5m\_vs\_6m     & {\underline{73}}    & {\underline{73}}    & \textbf{74}        & \textbf{74}         & 63                  \\
3s\_vs\_5z     & {\underline{94}}    & \textbf{97}         & 24                 & 92                  & 88                  \\
2c\_vs\_64zg   & {\underline{85}}    & \textbf{86}         & 21                 & 41                  & 64                  \\
MMM2           & \textbf{81}         & 5                   & 8                  & 1                   & {\underline{61}}    \\
bane\_vs\_bane & \textbf{99}         & 26                  & {\underline{96}}   & 81                  & \textbf{99}         \\
27m\_vs\_30m   & \textbf{71}         & 13                  & 13                 & 16                  & {\underline{30}}    \\ \hline
\end{tabular}
\label{tab:result}
\end{table}

\subsection{Visualization}
To visualize the latent states learned with SIDE, the t-SNE~\cite{Maaten2008VisualizingDU} plots are given in Figure~\ref{fig:vis}. We compare the embedding of real states with that of latent states on \emph{2c\_vs\_64zg} scenario after training has finished. For the real states, neighboring points in the embedding space often only indicate that the states they represent may belong to the same multi-agent decision-making process such as the points in the red circle, although their state values may be very different. In other words, there is no intuitive relationship between real states and state values, so it is not conducive to learning. On the contrary, points with the similar state values tend to cluster together in the embedding of latent states learned with SIDE. From the visualization, we can draw the conclusion that SIDE can infer the latent states that can accelerate the reinforcement learning process and we also understand why SIDE still has excellent performance even if global states are not available on \emph{2c\_vs\_64zg} scenario.


\section{Conclusion and Future work}
In this paper, in order to infer the current state based on the past local observation, we propose SIDE, a novel algorithm that combines multi-agent value decomposition and variational inference, so that no real state information is required during the entire training and execution process. Meanwhile, by jointly optimizing the two tasks of variational inference and optimal control, SIDE can promote the reconstruction of the state that is conducive to reinforcement learning, so that it far exceeds QMIX-PO, QMIX-HO and VDN in many tasks of SMAC, and even exceeds vanilla QMIX and other state-of-the-art methods.

SIDE can be easily extended to any multi-agent value decomposition algorithm with a mixing network, and the performances will be submitted in our future work. In addition, finding the interpretation of the reconstructed state is also a valuable research field, which will help enhance the robustness of multi-agent decision-making and help humans discover some unknown knowledge. It will also be the focus of our future work.


\bibliographystyle{ACM-Reference-Format} 


\onecolumn
\appendix
\numberwithin{equation}{section}
\numberwithin{figure}{section}
\numberwithin{table}{section}
\renewcommand{\thesection}{{\Alph{section}}}
\renewcommand{\thesubsection}{\Alph{section}.\arabic{subsection}}
\renewcommand{\thesubsubsection}{\Roman{section}.\arabic{subsection}.\arabic{subsubsection}}
\setcounter{secnumdepth}{-1}
\setcounter{secnumdepth}{3}

\section{Derivation of the ELBO of Dec-POMDPs}
The ELBO of Dec-POMDPs can be computed by the following equation:
\begin{align}
&\quad\log p\left(\mathcal{O}^{0: T}, \boldsymbol{u}^{0: T}, \boldsymbol{z}^{1: T}\right) \nonumber\\
&=\log \mathbb{E}_{q_{\theta}\left(s^{1: T} \mid \mathcal{O}^{1: T}, \boldsymbol{u}^{0: T}, \boldsymbol{z}^{1: T}\right)}\left[\frac{p\left(s^{1: T}, \mathcal{O}^{0: T}, \boldsymbol{u}^{0: T}, \boldsymbol{z}^{1: T}\right)}{q_{\theta}\left(s^{1: T} \mid \mathcal{O}^{0: T}, \boldsymbol{u}^{0: T}, \boldsymbol{z}^{1: T}\right)}\right] \nonumber\\
& \geq \mathbb{E}_{q_{\theta}\left(s^{1: T} \mid \mathcal{O}^{1: T}, \boldsymbol{u}^{0: T}, \boldsymbol{z}^{1: T}\right)} \log \left[\frac{p\left(s^{1: T}, \mathcal{O}^{0: T}, \boldsymbol{u}^{0: T}, \boldsymbol{z}^{1: T}\right)}{q_{\theta}\left(s^{1: T} \mid \mathcal{O}^{0: T}, \boldsymbol{u}^{0: T}, \boldsymbol{z}^{1: T}\right)}\right]\nonumber\\
&=\int q_{\theta}\left(s^{1: T} \mid \mathcal{O}^{1: T}, \boldsymbol{u}^{0: T}, \boldsymbol{z}^{1: T}\right) \log \left[\frac{p\left(s^{1: T}, \mathcal{O}^{0: T}, \boldsymbol{u}^{0: T}, \boldsymbol{z}^{1: T}\right)}{q_{\theta}\left(s^{1: T} \mid \mathcal{O}^{0: T}, \boldsymbol{u}^{0: T}, \boldsymbol{z}^{1: T}\right)}\right] d s^{1: T}\nonumber \\
&=\int \sum_{t=1}^{T} q_{\theta}\left(s^{1: T} \mid \mathcal{O}^{1: T}, \boldsymbol{u}^{0: T}, \boldsymbol{z}^{1: T}\right) \log \left[\frac{p\left(\boldsymbol{u}^{t}\right) p\left(\mathcal{O}^{t} \mid s^{t}, \boldsymbol{u}^{t}\right) p\left(s^{t} \mid s^{t-1}, \boldsymbol{u}^{t-1}\right) p\left(\boldsymbol{z}^{t} \mid s^{t}\right)}{q_{\theta}\left(s^{t} \mid s^{t-1}, \boldsymbol{u}^{t-1}, \boldsymbol{z}^{t}\right)}\right] d s^{1: T} \nonumber\\
&=\sum_{t=1}^{T} \int q_{\theta}\left(s^{1: t} \mid \mathcal{O}^{1: t}, \boldsymbol{u}^{0: t}, \boldsymbol{z}^{1: t}\right) \log \left[\frac{p\left(\boldsymbol{u}^{t}\right) p\left(\mathcal{O}^{t} \mid s^{t}, \boldsymbol{u}^{t}\right) p\left(s^{t} \mid s^{t-1}, \boldsymbol{u}^{t-1}\right) p\left(\boldsymbol{z}^{t} \mid s^{t}\right)}{q_{\theta}\left(s^{t} \mid s^{t-1}, \boldsymbol{u}^{t-1}, \boldsymbol{z}^{t}\right)}\right] d s^{1: t} \nonumber\\
&=\sum_{t=1}^{T}\left\{\int q_{\theta}\left(s^{1: t} \mid \mathcal{O}^{1: t}, \boldsymbol{u}^{0: t}, \boldsymbol{z}^{1: t}\right) \log \left[p\left(\boldsymbol{u}^{t}\right) p\left(\mathcal{O}^{t} \mid s^{t}, \boldsymbol{u}^{t}\right) p\left(\boldsymbol{z}^{t} \mid s^{t}\right)\right] d s^{1: t}\right. \nonumber\\
&\quad \left.+\int q_{\theta}\left(s^{1: t} \mid \mathcal{O}^{1: t}, \boldsymbol{u}^{0: t}, \boldsymbol{z}^{1: t}\right) \log \left[\frac{p\left(s^{t} \mid s^{t-1}, \boldsymbol{u}^{t-1}\right)}{q_{\theta}\left(s^{t} \mid s^{t-1}, \boldsymbol{u}^{t-1}, \boldsymbol{z}^{t}\right)}\right] d s^{1: t}\right\}\nonumber\\
&= \sum_{t=1}^{T}\left\{\int q_{\theta}\left(s^{1: t} \mid \mathcal{O}^{1: t}, \boldsymbol{u}^{0: t}, \boldsymbol{z}^{1: t}\right) \log \left[p\left(\boldsymbol{u}^{t}\right) p\left(\mathcal{O}^{t} \mid s^{t}, \boldsymbol{u}^{t}\right) p\left(\boldsymbol{z}^{t} \mid s^{t}\right)\right] d s^{1: t}\right.\nonumber\\ 
&\left.\quad -\int q_{\theta}\left(s^{1: t-1} \mid \mathcal{O}^{1: t-1}, \boldsymbol{u}^{0: t-1}, \boldsymbol{z}^{1: t-1}\right) \mathcal{D}_{K L}\left[q_{\theta}\left(s^{t} \mid s^{t-1}, \boldsymbol{u}^{t-1}, \boldsymbol{z}^{t}\right) \| p\left(s^{t} \mid s^{t-1}, \boldsymbol{u}^{t-1}\right)\right] d s^{1: t}\right\} \nonumber\\
&= \mathbb{E}_{q_{\theta}\left(s^{1: T} \mid \mathcal{O}^{1: T}, \boldsymbol{u}^{ 0: T}, \boldsymbol{z}^{1: T}\right)} \sum_{t=1}^{T}\left\{\log \left[p\left(\boldsymbol{u}^{t}\right) p\left(\mathcal{O}^{t} \mid s^{t}, \boldsymbol{u}^{t}\right) p\left(\boldsymbol{z}^{t} \mid s^{t}\right)\right] \right.\nonumber\\ 
&\left.\quad-\mathcal{D}_{K L}\left[q_{\theta}\left(s^{t} \mid s^{t-1}, \boldsymbol{u}^{t-1}, \boldsymbol{z}^{t}\right) \| p\left(s^{t} \mid s^{t-1}, \boldsymbol{u}^{t-1}\right)\right]\right\} \nonumber\\ 
&= \sum_{t=1}^{T}\left\{r\left(s^{t}, \boldsymbol{u}^{t}\right)+\log \left[p\left(\boldsymbol{u}^{t}\right) p\left(\boldsymbol{z}^{t} \mid s^{t}\right)\right]\right.\nonumber\\ 
&\quad -\mathcal{D}_{K L}\left[q_{\theta}\left(s^{t} \mid s^{t-1}, \boldsymbol{u}^{t-1}, \left.\boldsymbol{z}^{t}\right) \| p\left(s^{t} \mid s^{t-1}, \boldsymbol{u}^{t-1}\right)\right]\right\}\label{eq:lasttwo}\\
&=\sum_{t=1}^{T}\left\{r\left(s^{t}, \boldsymbol{u}^{t}\right)+\sum_{a=1}^n\log \left[p\left(u_a^{t}\right)\right] + \sum_{a=1}^n\log \left[p\left(z_a^{t} \mid s^{t}\right)\right]\right.\nonumber\\
&\left.\quad-\mathcal{D}_{K L}\left[q_{\theta}\left(s^{t} \mid s^{t-1}, \boldsymbol{u}^{t-1}, \boldsymbol{z}^{t}\right) \| p\left(s^{t} \mid s^{t-1}, \boldsymbol{u}^{t-1}\right)\right]\right\},\label{eq:lastone}
\end{align}
where $s^{1: T} \sim q_{\theta}\left(s^{1: T} \mid \mathcal{O}^{1: T}, \boldsymbol{u}^{0: T}, \boldsymbol{z}^{1: T}\right)$. Equation~\ref{eq:lasttwo} can be written as Equation~\ref{eq:lastone} under the conditions as follows:
\begin{equation*}
\begin{aligned}
    p(\boldsymbol{u^t}) &= \prod_{a = 1}^n p(u_a^t), \\
    p(\boldsymbol{z}^t\mid s^t) &= \prod_{a=1}^n p(z_a^t\mid s^t).
\end{aligned}
\end{equation*}
Even if these assumptions are not true in most multi-agent systems, we verified that SIDE can achieve excellent performance through many experiments.

\section{Experiment Detail}
Table~\ref{tab:hyperparameters} summarizes hyperparameters used in our implementation.

\begin{table}[ht]
\caption{The hyperparameters setup.}
\begin{tabular}{ll}
\hline
\textbf{Description}                                        & \textbf{Value} \\ \hline
Learning rate                                               & 0.0005         \\
Type of optimizer                                           & RMSProp        \\
RMSProp param  $\alpha$                                     & 0.99           \\
RMSProp param  $\epsilon$                                   & 0.00001        \\
How many episodes to update target networks                 & 200            \\
Reduce global norm of gradients                             & 10             \\
Batch size                                                  & 32             \\
Capacity of replay buffer (in episodes)                     & 5000           \\
Discount factor  $\gamma$                                   & 0.99           \\
Exploration rate annealing time (in timesteps)              & 50000          \\
Starting value for exploraton rate annealing                & 1              \\
Ending value for exploraton rate annealing                  & 0.05           \\
How many timesteps to run                                   & 2050000        \\
Size of hidden state for the rnn agent                      & 64             \\
Size of embedding layer for the mixing net                  & 32             \\ \hline
Size of latent dimension for each agent  $N$                & 64             \\
Size of embedding layer for the prior function and the VGAE & 128            \\ \hline
\end{tabular}
\label{tab:hyperparameters}
\end{table}

\section{Additional Experiments with other SOTA methods}

\begin{figure*}[ht]
    \centering
    \includegraphics[width= 0.95 \textwidth]{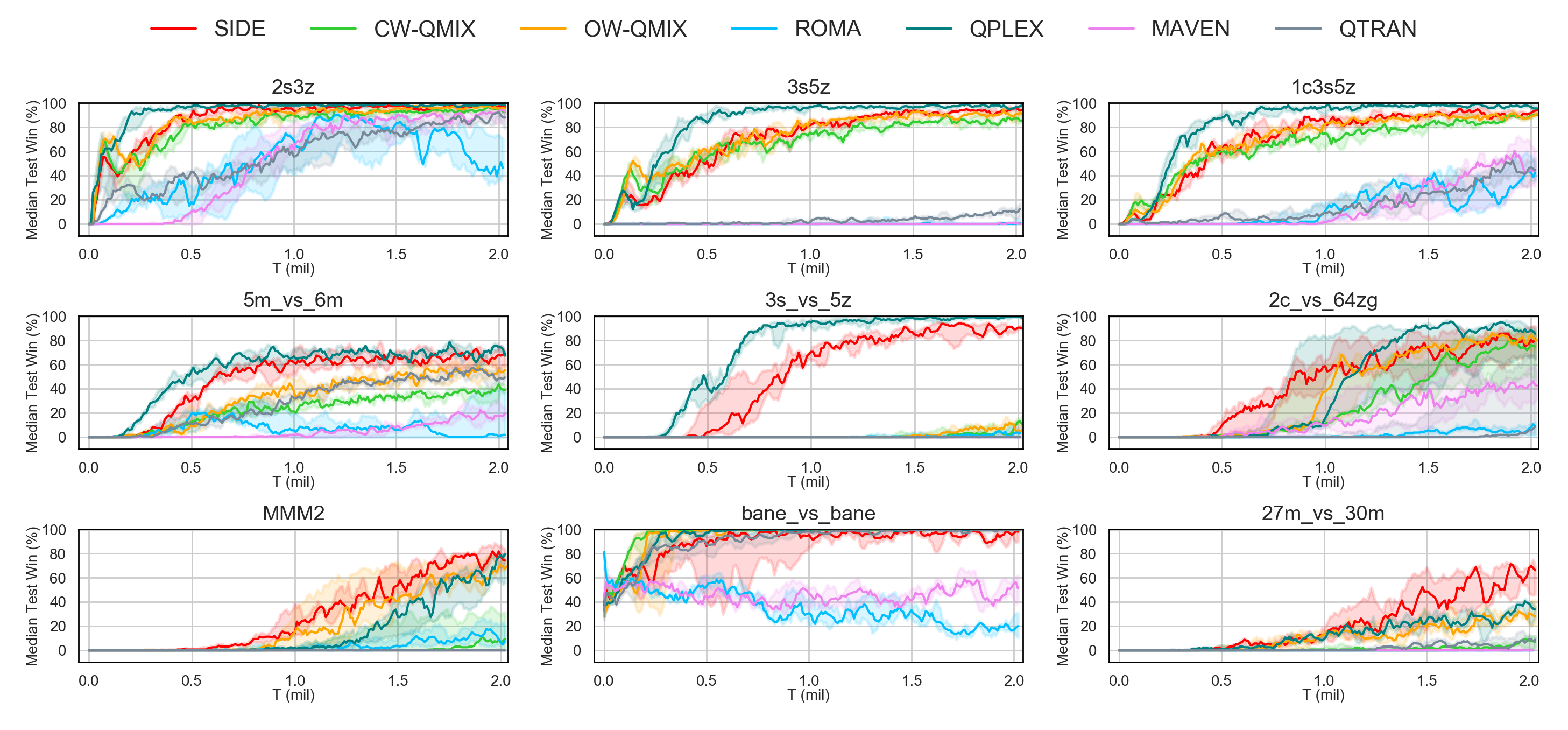}
    \caption{Overall results in different scenarios with current popular baselines. Note that only SIDE cannot get the global state.}
    \label{fig:result_sota}
\end{figure*}

For evaluation, we compare with some popular value decomposition methods and provide more experimental results. The results are depicted in Figure~\ref{fig:result_sota}. The experiment setup is same as the one mentioned previously. Noted that the results of some baselines are different from that in the literature and we think it may be caused by the version mismatch. The version of StarCraft II we used in this paper is 4.6.2 (B69232), which is more difficult than 4.10. From this figure it can be seen that SIDE has the competitive performance although SIDE cannot access the state. So we can draw the conclusion that SIDE can infer the latent state which is beneficial to the reinforcement learning process. Besides, we also look forward to the performance of SIDE based on other value-based algorithms, such like QPLEX and so on.

\end{document}